\documentstyle[prb,aps,epsfig,multicol]{revtex}

\begin{document}
\draft
\title{Effect of sintering temperature under high pressure in the superconductivity
for MgB$_{2}$}
\author{C. U. Jung,\cite{email} Min-Seok Park, W. N. Kang, Mun-Seog Kim, Kijoon H.
P. Kim, S. Y. Lee, and Sung-Ik Lee}
\address{National Creative Research Initiative Center for Superconductivity\\
and Department of Physics, Pohang University of Science and Technology,\\
Pohang 790-784, Republic of Korea \\
}
\date{\today }
\maketitle

\begin{abstract}
We report the effect of the sintering temperature on the superconductivity
of MgB$_{2}$\ pellets prepared under a high pressure of 3 GPa. The
superconducting properties of the non-heated MgB$_{2}$ in this high pressure
were poor. However, as the sintering temperature increased, the
superconducting properties were vastly enhanced, which was shown by the
narrow transition width for the resistivity and the low-field
magnetizations. This shows that heat treatment under high pressure is
essential to improve superconducting properties. These changes were found to
be closely related to changes in the surface morphology observed using
scanning electron microscopy.
\end{abstract}



\begin{multicols}{2}

 The recent discovery of superconductivity at about 40 K in
MgB$_{2}$\ has resulted in a surge of interest.\cite{Akimitsu}
Conventional BCS
superconductivity has been proposed for this compound, and a shift in the $%
T_{c}$ due to the boron isotope has been reported with an isotope critical
exponent of $\alpha _{B}\sim 0.26$.\cite{Budko,Kortus} The type of carrier
has been predicted to be positive with boron planes acting like the CuO$_{2}$
planes in cuprate high-temperature superconductors.\cite{Hirsch} We
demonstrated a hole-type carrier for MgB$_{2}$ by using Hall measurements.
\cite{WNKang} Note that other transition metal borides, which are not
superconducting, have been shown to have negative Hall coefficients.\cite
{Juretschke,Johnson}

Previously MgB$_{2}$ was synthesized from a stoichiometric mixture of Mg and
B in a sealed Ta tube which was placed in a quartz ampoule at $950^{\circ }$%
C.\cite{Budko,Finnemore,Paul} However, these samples were found to be porous
and mechanically weak.\cite{Finnemore,Paul} Some have reported the synthesis
of dense pellets sintered under high pressures of several GPa,\cite
{Jung,Ihara,Cava} which makes it easy to study the transport properties. The
physical properties for samples made at different pressures have been
different from each other,\cite{Jung,Ihara} \ yet so far, the effect of the
synthesis conditions on the superconductivity \ has not been studied.

Now the binary intermetallic compound MgB$_{2}$ has become a strong
candidate for superconductivity applications because of its metallic nature
\ and its simple structure compared to oxide superconductors. Moreover, the
compound is expected to be less anisotropic than layered high-temperature
superconductors. However, the growth of a thin film has not yet been
reported, the high melting temperature of born being one of the major
reasons. A proper target for synthesizing a thin film has not yet been
fabricated. The most promising candidate is pulsed laser ablation technique,
but large dense target is not yet available.\cite{densetarget}

For large-scale device applications, as well as for small-scale
electronic device applications, the relation between the synthesis
conditions and the superconducting properties should be well
established. Previously, we
reported the transport properties of hard and dense MgB$_{2}$ sintered at $%
950^{\circ }$C under 3 GPa. We found a clear difference in the resistivity
as a function of the temperature and the magnetic field between our sample
prepared at high pressure and other bulk samples not prepared at high
pressure.\cite{Jung} The metallic nature of our sample could be inferred
from its shiny surface. This sample was strong enough to prepare an
optically clean and flat surface for a reflectivity measurement after
polishing.

In this paper we report the change in the resistivity near $T_{c}$, the
low-field magnetization, and the surface morphology of MgB$_{2}$ with the
sintering temperature at 3 GPa. As the sintering temperature of the MgB$_{2}$
pellet was increased from 500 to $950^{\circ }$C, the superconducting
transition width for a 10 to 90\% drop, $\Delta T_{c}$, decreased
systematically while the onset temperature, $T_{co}$, was nearly the same.
Scanning electron microscope images showed that these behaviors were
accompanied by changes in the grain size and their connectivity.

A 12-mm cubic multi-anvil-type press was used for the high-pressure
sintering.\cite{JungLa} The starting material was a commercially available
powder of MgB$_{2}$.\cite{Alfa} The pressed pellet was put into a Au capsule
in a high-pressure cell and pressurized up to $3$ GPa. While the pressure
was maintained, the heating was increased linearly; then, maintained at
constant temperature for 2 hours. The samples were then quenched to room
temperature. The S(500), S(700), S(800), and S(950) in this paper were
pellets sintered at 500, 700, 800, and 950$^{\circ }$C,\ respectively. The
pellets, weighing about 130 mg, were about 4.5 mm in diameter and 3.3 mm in
height. For samples sintered at temperatures higher than $950^{\circ }$C,
the gold started to melt and to adhere strongly to the MgB$_{2}$.

A dc SQUID magnetometer (Quantum Design, MPMS{\it XL}) and a
field-emission scanning electron microscope (SEM) were used to
investigate the low-field magnetization and the surface
morphology. For the resistivity measurement, we cut the samples by
using a diamond saw and then polished them into
rectangular solid shapes with dimensions of about $1\times 1\times 4.5$ mm$%
^{3}.$ The resistivity curve, $R$($T$), was measured by using a
standard 4-probe technique.

\begin{figure}
\centering
 \epsfig{file=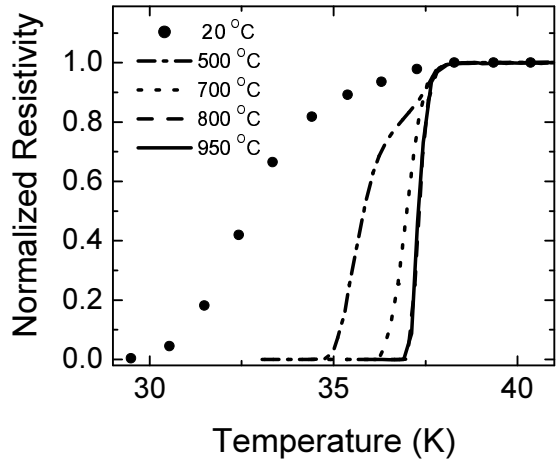, width=7.5cm}
\caption{Normalized resistivity of sintered MgB$_{2}$ made under 3
GPa. The resistivity values were normalized to 1.0 at 40 K, to
observe the change in the superconducting transition width with
the sintering temperature. The solid circles denoted by 20$^{\circ
}$C\ are for the pellet which was pressurized without actual
heating. The solid, dashed, dotted, and
dash-dotted line are for the pellet sintered at 950, 800, 700, and 500$%
^{\circ }$C, respectively.} \label{RT}
\end{figure}

Figure \ref{RT} shows the normalized resistivity of the sintered pellets
near $T_{c}$. The resistivity values were normalized to R(40\ K), just above $%
T_{c}$, for comparison. The solid circles represent the
resistivity for the S(20) which was pressurized at 3 GPa without
subsequent heating for a conducting current path. The solid,
dashed, dotted, and dash-dotted lines are for S(950), S(800),
S(700), and S(500), respectively.

As the sintering temperatures was increased from 500 to 800 $^{\circ }$C,
the $\Delta T_{c}$\ decreased systematically, but the $T_{co}$ changed
little. Also, the $\Delta T_{c}$ of S(800) and S(950) were nearly the same,
which indicates that the optimum sintering temperature region is wide under
high pressure.\cite{Finnemore} The absolute value of the resistivity at 40 K
of S(950) was nearly the same as that of S(800), but was about two times
smaller than that of S(500).

\begin{figure}
\centering
 \epsfig{file=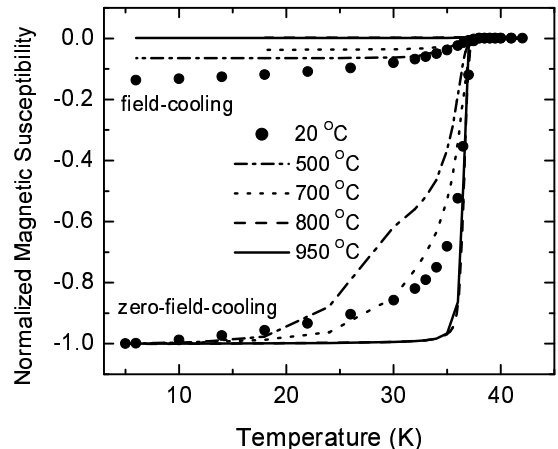, width=7.5cm}
\caption{Normalized magnetic susceptibility from the low-field
magnetization $M(T)$\ of sintered MgB$_{2}$ made under 3 GPa. The
solid circles are for the as-purchased powder. Without these data,
the zero-field-cooling lines
from the top correspond to the samples sintered at 500, 700, 800, and 950$%
^{\circ }$C, respectively. The curves for S(800) and S(950) nearly
overlay each other. The pronounced hump around 30 K shown for
S(500) seems to be due to a weak-link.} \label{lowfieldMT}
\end{figure}

Figure \ref{lowfieldMT} shows the normalized magnetic
susceptibility, $4\pi \chi (T)$, from the measured\ low-field
magnetization of MgB$_{2}$. The solid circles are for the starting
powder itself. The solid, dashed, dotted, and dash-dotted lines
are for S(950), S(800), S(700), and S(500), respectively. The
lines for S(950) and S(800) nearly overlay each other. The
pronounced hump around 30 K shown for S(500) \ was due to a
weak-link. This weak-link feature was also visible in the
resistivity data in Fig. \ref{RT}. As the sintering temperature
was increased above about 700$^{\circ }$C, the weak-line feature
disappeared. These support that, in high pressure,
originally-separated grains begin to develope a weak-link forms below 500$%
^{\circ }$C, and gradually a strong-link. This trend will be shown
in the following paragraphs. The decrease in superconducting
transition width is more evident in the $\chi _{ZFC}(T)$
measurement than in the resistivity measurement. The field-cooling
signal was found to mirror the $\chi _{ZFC}(T) $; namely the
pinning seems to be enhanced as the sintering temperature is
increased. For samples sintered above 800$^{\circ }$C, the
field-cooling signal was less than 0.5 percent of the
zero-field-cooling signal.

\begin{figure}
\centering
 \epsfig{file=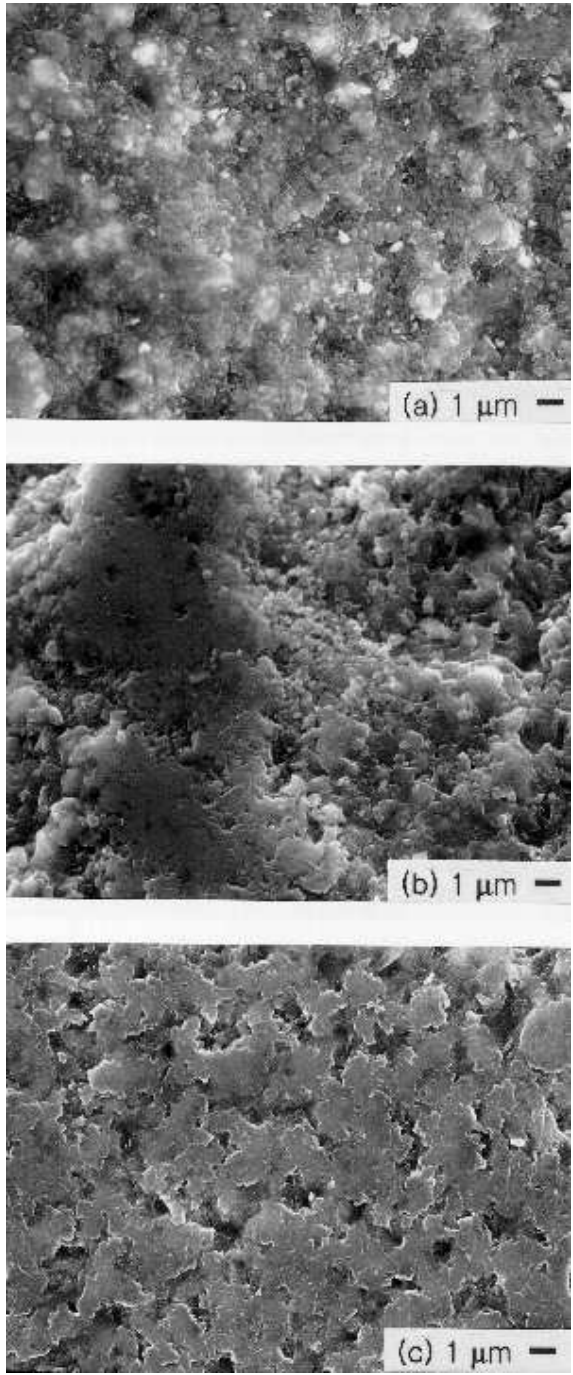, width=7.5cm}
\caption{SEM pictures of sintered MgB$_{2}$ made under 3 GPa. The
scale bars are 1 $\protect\mu m$ in length. The sintering temperatures were (a) 20$%
^{\circ }$C, (b) 500$^{\circ }$C, and (c) 950$^{\circ }$C. The
sample without actual heating, S(20), shows well-separated grains
with spacious voids. As the sintering temperature is increased,
the connectivity of each grain increases and the porosity
decreases rapidly. } \label{SEM}
\end{figure}

Figure \ref{SEM} shows SEM images of S(20), S(500), and S(950) with a
magnification of 5,000. The scale bars are 1 $\mu $m in length. The image in
Fig. \ref{SEM} (a) is for S(20). Here, we can clearly see separated grains
and voids. The grain size is less than about $0.5$ $\mu $m. In Fig. \ref{SEM}
(b), some parts of the grains for S(500) adhered to each other while other
parts of the grains remain isolated. In Fig. \ref{SEM} (c), the S(950)
pellet has grains that are very well connected with each other. We cannot
even distinguish one grain from another over a wide area. Energy dispersive
spectroscopy using SEM showed that gold used to wrap around the pellet
smeared into the pellet by a depth of about 300 $\mu $m for S(950). This
feature was very helpful for attaching electric pads and could be another
small advantage of this material for electrical applications.

We observed well-interconnected changes in the transport and the magnetic
properties, near $T_{c}$, which were accompanied by the microscopic changes
in the observed surface morphology of MgB$_{2}$. The strong connection
between the grains with increasing sintering temperature seems to explain
the change in the resistivity and the low-field magnetization.

We are thankful for the SEM work to Mr. D. S. Kim at the Department of
Material Science and Engineering. Discussion with Dong-Wook Kim on the film
growth is also acknowledged. This work is supported by the Ministry of
Science and Technology of Korea through the Creative Research Initiative
Program.

\end{multicols}

\end{document}